\documentclass[11pt]{article}
\usepackage{amsmath,amsfonts, amssymb,graphicx,geometry,authblk,color,soul,comment,bm,physics,subcaption} 
\usepackage{cleveref}
\title{High Strain Rate Behavior of Liquid Crystal Elastomers}
\author[1]{Adeline Wihardja}
\author[2]{Juan Carlos Nieto Fuentes}
\author[3]{Daniel Rittel}
\author[1]{Kaushik Bhattacharya }
\affil[1]{Department of Mechanical and Civil Engineering, California Institute of Technology, USA}
\affil[2]{Department of Mechanical Engineering, Rey Juan Carlos University, Calle Tulipan, S/N-28933-Mostoles, Spain}
\affil[3]{Faculty of Mechanical Engineering, Technion -- Israel Institute of Technology, Israel}

\date{}

\begin{document}
\maketitle

\begin{abstract}
Liquid crystal elastomers are rubbery solids that couple liquid crystalline order and deformation.  This coupling leads to properties that are attractive for a number of applications in soft robotics and energy absorption.  This paper is motivated by the latter application, and provides a systematic experimental study of a particular class of liquid crystal elastomers -- the isotropic genesis polydomain liquid crystal elastomers -- over a wide range of strain rates.  An important aspect of this study is a novel tensile drop-tower that enables tensile strain rates of 100 s$^{-1}$ that are important to application but previously inaccessible.  The paper also extends a recently proposed constitutive model to the high strain rate regime, and shows that it can be fit to describe the observed behavior across the spectrum of examined behavior.

\end{abstract}

\section{Introduction} \label{sec:intro}

Liquid crystal elastomers (LCEs) are cross-linked elastomers in which stiff nematic mesogens are incorporated into the polymer chains \cite{warner_book,terentjev_2025}.   The cross-linking is light enough to enable the mesogens to undergo order-disorder transitions.  In nematic elastomers, the mesogens are randomly aligned in a disordered phase or isotropic phase above a transition temperature, but align around a particular direction (director) in the nematic phase below the transition temperature. Since these mesogens are incorporated in the polymer chains, the isotropic to nematic transition leads to significant spontaneous or residual elongation along the director.   The transition can also be triggered by other stimuli, including solvents, electric field, magnetic field, and light, depending on the chemistry.  This interplay between temperature, nematic order,  deformation, and stimuli makes LCE a promising material for actuation \cite{palagi_actuation, zeng_actuation, yuan_actuation}, mechanical dissipation \cite{clarke_2001, Traugutt,saed_impact}, and energy absorption \cite{jeon_impact, saed_impact}. 

Various types of LCEs exist depending on their chemistry and synthesis history.   There are both main-chain \cite{bergmann,yakacki_RSC} and side-chain \cite{finkelmann} LCEs.  Further, they can be synthesized in both the monodomain configuration (MLCE) \cite{kupfer_finkelmann}, where all mesogens are aligned uniformly during synthesis, or in the polydomain configuration (PLCE), where the mesogens form domains each having its own local orientation during synthesis.    Since PLCEs can be made without requiring any alignment, they are scalable and easy to manufacture, making them good engineering materials \cite{yakacki_RSC}. There are two main types of PLCEs: isotropic-genesis PLCEs (I-PLCEs) and nematic-genesis PLCEs (N-PLCEs). I-PLCEs are made via cross-linking in the high-symmetry isotropic phase in solution before removing the solvent to trigger the phase transition to the nematic phase.  I-PLCEs typically have a large number of randomly oriented domains in the nematic phase, and the material retains its overall isotropy.  N-PLCEs are cross-linked in the nematic phase. Importantly, this different cross-linking state gives rise to dissimilar mechanical behavior, especially concerning the soft behavior \cite{urayama_kohmon,biggins_warner_bhatta}. 

A unique feature of LCEs is their soft behavior, where they can undergo large stretch under low, constant stress \cite{warner_book, kupfer_finkelmann,clarke_1998,fridrikh_terentjev_1999,urayama_kohmon}.  While initially discovered in MLCEs \cite{kupfer_finkelmann}, the soft behavior is also observed in I-PLCEs \cite{clarke_1998,fridrikh_terentjev_1999,urayama_kohmon} as a result of a polydomain to monodomain transition \cite{fridrikh_terentjev_1999,clarke_1998,biggins_warner_bhatta,zhou}.  Indeed, this soft behavior can manifest itself in rather complex ways, including an in-plane liquid-like behavior, i.e., they are unable to sustain shear stresses in the plane \cite{tokumoto_zhou}.   

This soft behavior is known to be extremely sensitive to the strain rate \cite{hotta_terentjev, clarke_1998, azoug, linares}.  This indicates that the polydomain to monodomain transition, and mesogen reorientation more broadly, is rate-dependent.  This is supported by the fact that these materials have a large loss tangent over a wide range of temperature and frequencies (75-110$^{\circ}\text{C}$, 1-100 Hz) \cite{clarke_2001,Traugutt}.

The combination of soft behavior and large mechanical damping makes these materials extremely attractive for energy absorption and mechanical protection applications.  For example, Jeon \textit{et al.} \cite{jeon_impact} showed that I-PLCE-based architected material displays extremely large energy absorption density (5MJ/m$^3$) at a high overall strain rate of 600 s$^{-1}$.   This has motivated the study of the rate-dependent behavior of I-PLCEs.  These include small amplitude vibration characterization using a Dynamic Mechanical Analysis, quasi-static conditions in the range of $10^{-4} - 10^{-1}$ s$^{-1}$ using universal testing machines, and compression tests at very high strain rates of 10$^3$ s$^{-1}$ using a split Hopkinson pressure bar (SHPB) or Kolsky bar \cite{saed_impact}.   Unfortunately, this leaves a critical gap at strain rates of $10^2$ s$^{-1}$, especially in tension, where much of the interesting physics of impact mitigation and soft behavior occurs.   

This paper addresses this critical gap in thiol-acrylate-based main chain I-PLCEs.  We conduct quasistatic tests in both tension and compression using a universal testing machine, dynamic tension tests at strain rates of strain rates of $10^2$ s$^{-1}$ using a newly developed tensile testing apparatus \cite{nieto2025new}, and dynamic compression tests at strain rates of strain rates of $10^3$ s$^{-1}$ using a split Hopkinson pressure bar.  Together, they provide a comprehensive description of the strain-rate behavior of the I-PLCEs.  

We also develop a macroscopic model of this complex strain-rate-dependent soft behavior by extending the model of Lee {\it et al.} \cite{lee_2023} to include rate-dependence, viscoelasticity and compressibility.   I-PLCEs have multiscale nature as has been discussed elsewhere \cite{urayama_kohmon,biggins_warner_bhatta,zhou,lee_2023}.  We are interested in the macroscopic or engineering scale where a typical representative volume consists of a large number of domains.   Each domain has a relatively uniform director, i.e., the nematic mesogens are aligned almost uniformly about a particular direction or director.  The mechanical behavior of the elastomer at the domain scale is described by the celebrated Bladon-Warner-Terentjev theory \cite{blandon_1993}.  This theory is characterized by an {\it anisotropy parameter} $r$ that describes the ratio of the spontaneous or residual stretch in the direction of the director to that perpendicular to the director.  This theory is non-convex and leads to domains in monodomain LCEs, and the relaxation of this theory has been computed by Desimone and Dolzmann \cite{desimone_dolzmann}.  In I-PLCEs, one has domains  due to relaxation, but also due to synthesis, and thus one needs a combination of relaxation and homogenization.  This is studied with bounds in Biggins {\it et al.} \cite{biggins_warner_bhatta} and with numerical simulation in Zhou and Bhattacharya \cite{zhou}.  Importantly, the domain patterns as well as the orientation of the directors within the domain, and consequently the spontaneous deformation, can change when the I-PLCE is subject to stress.  These results are the basis of the macroscopic constitutive model of Lee {\it et al.} \cite{lee_2023}.  The key idea of this theory is that the spontaneous deformation of I-PLCEs is characterized by two internal variables  $\Lambda$ and $\Delta$, and the behavior is characterized by the evolution of these internal variables.  This theory is rich enough to address monodomain and polydomain LCEs.  The strain rate response of LCEs depends on both the kinetics of evolution and the viscoelasticity of the polymer network.  This has been modeled by Wang {\it et al.} \cite{govindjee} focusing on monodomain LCEs, building on the viscoelastic framework of 
Reese and Govindjee \cite{reese1998theory}.  In this work, we extend the model of Lee {\it et al.} \cite{lee_2023} to include network viscoelasticity. 
We show that all our experimental observations can be fit to a single set of parameters in this model, and that the model provides further insight into the polydomain-monodomain transition at various rates.


\section{Experimental Methods} \label{sec:exp}

\subsection {Liquid Crystal Elastomers Synthesis} \label{sec:baking}

We study main chain, thiol-acrylate-based, isotropic-genesis polydomain LCEs (I-PLCEs)  fabricated using Michael addition click reaction \cite{yakacki_RSC} (also \cite{lee_2021}).


\subsection{Tension tests} \label{sec:MTS}

\begin{figure}[t]
\centering
\includegraphics[width=0.65\textwidth]{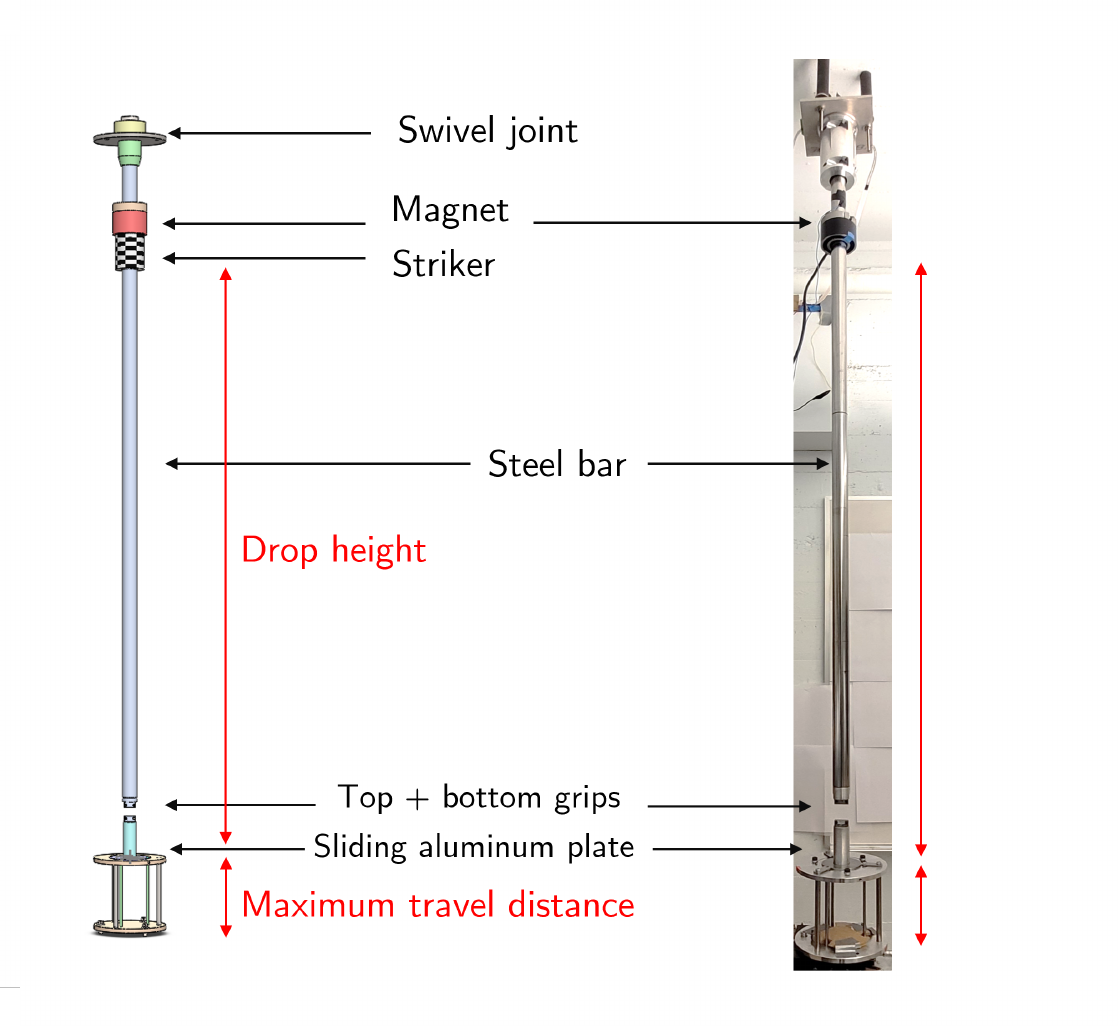}
\caption{Schematic drawing (left) and image (right) of the drop weight tensile testing system. Note that the striker is not shown in the right image.}
\label{fig:droptower}
\end{figure}

Quasistatic tests, at a strain rate of $\dot{\varepsilon} = 10^{-2}$ s$^{-1}$, are conducted using a universal testing machine, MTS Model 358.10. 

Intermediate strain rate tests are conducted with a custom-made drop tower setup shown in Figure \ref{fig:droptower}, and discussed in detail in Nieto-Fuentes \textit{et al.} \cite{nieto2025new}. Briefly, the setup consists of a stainless steel striker, which is guided down by a stainless steel bar. At the bottom of this bar is a clamp that secures one edge of the I-PLCE specimen. The other edge of the specimen is clamped to the bottom clamp, which is connected to a sliding aluminum plate capable of traveling vertically up and down. When the striker is released, it travels through the guiding rod until it hits the bottom clamp, pushing it downward and applying a tensile force on the specimen. The impact velocity and strain rate of the material can be controlled by adjusting the drop height of the striker. 

A series of images is taken immediately following impact, using a high-speed camera (Photron Fastcam NOVA S12 camera and Tokina AT-X PRO lens, 100F 2.8D) at 10,000 frames per second with a resolution of 1024 by 1024 pixels under a continuous light source.   The images are used to obtain the displacement field within the specimen and the displacement of the bottom clamp using 2D digital image correlation (DIC) with the Correlated Solutions Vic2D software.   The subset size is 21$\times$21 pixels and the step size is 1 pixel. The Green-Lagrange strain tensor ($E = \frac{1}{2}(F^T F-I)$ where $F$ is the deformation gradient) is calculated using a strain filter of 5. The forces exerted by the specimen during impact are measured using two force sensors (Tekscan FlexiForce\texttrademark A201 of 445 N limit) at the top of the setup. After the striker is launched, once the stress wave reaches the force sensors, diagnostics are triggered at 200 MHz, and a Siglent SDS 1202X-E digital oscilloscope records the force data. 

All tension tests are carried out with a specimen shown in Figure \ref{fig:typical}(a).

\subsection{Compression tests} \label{sec:shpbsetup}

\begin{figure}
\centering
    \includegraphics[width=0.8\textwidth]{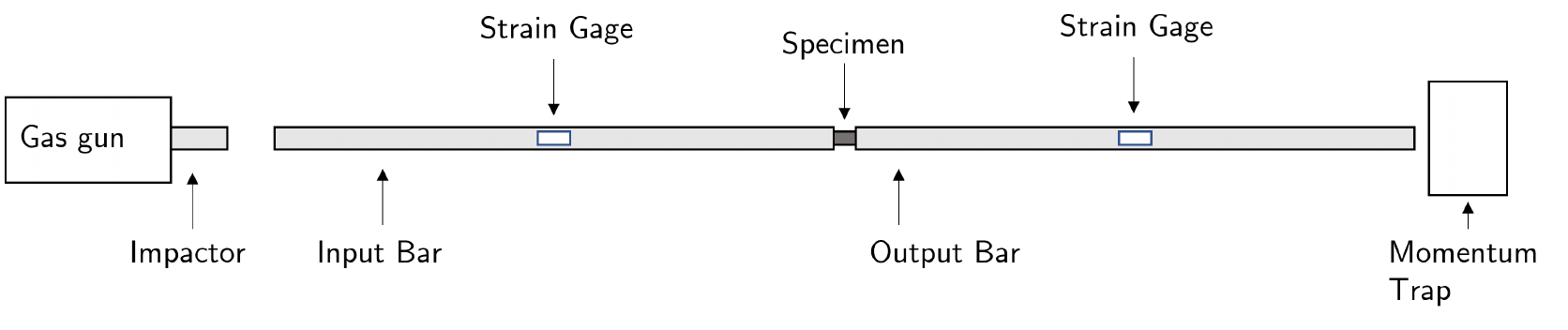}
    \caption{Schematic drawing of the split Hopkinson Pressure Bar (SHPB) setup consisting of impactor, input bar, and output bar}
    \label{fig:shpb}
\end{figure}

Quasistatic compression tests are conducted on I-PLCE pads with diameter of 10 mm and thickness of 3 mm, at a strain rate of $\dot{\varepsilon} = 10^{-2}$ s$^{-1}$, using a universal testing machine, MTS Model 358.10. 

High strain rate compression tests are conducted on I-PLCE pads of diameter 6.7 mm and thickness 3 mm using a split-Hopkinson pressure bar (SHPB) detailed in \cite{chen2010split} and shown schematically in Figure \ref{fig:shpb}.  The setup consists of an impactor, an incident bar, and an output bar. Strain gages (Omega SGD-2D/350-LY1) are placed in the middle of the incident and output bars, and the specimen is placed between the input and output bars.  Petroleum jelly is used between the specimen and the bars to minimize the friction between the specimen and the bars.  The strain gages are connected to a signal conditioning amplifier (Vishay 2310B), which connects to the digital oscilloscope (2.5 GHz Tektronix DPO 3014).  A Maraging 350 steel (AMS 6515) is used for the impactor (20/30 cm), incident bar (1.24 m), and output bar (1.24 m). The impactor, launched into the system using a gas gun, hits one end of the input bar, giving rise to a compressive stress wave that travels through the input bar to the specimen. Once this stress wave reaches the incident strain gage, diagnostics are triggered, and the oscilloscope records the strain gage voltage data. 
A Photron Fastcam NOVA S12 camera with 100 mm Tokina AT-X Pro lens with a continuous light source is used once again to capture the images of the specimen during impact at 200,000 frames per second with a resolution of 256 $\times$ 128 pixels. \\

\section{ Constitutive Model of I-PLCE} \label{sec:model}

\subsection{Kinematics}

\begin{figure}
\centering
\includegraphics[width=2in]{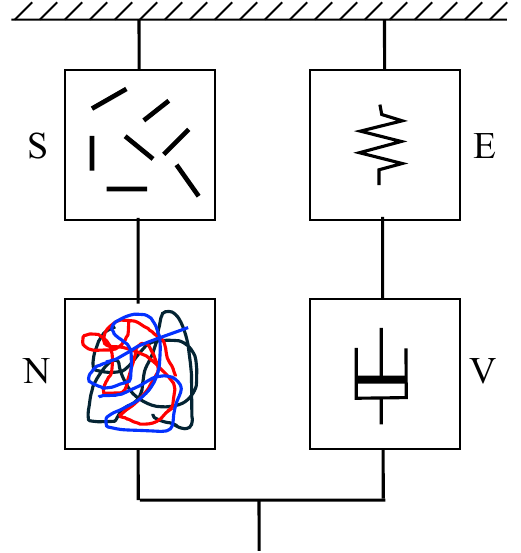}
\caption{Rheological model of LCE \label{fig:rheo}}
\end{figure}

We work at the macroscopic or engineering scale.  We consider a specimen of I-PLCE occupying a region $\Omega$ in its natural reference configuration at a temperature above the isotropic-nematic transition temperature $T_\text{ni}$.  We denote the deformation as $y: \Omega \to {\mathbb R}^3$, the deformation gradient as $F$, and the temperature by $T$.  

We model the I-PLCE using the rheological model shown schematically in Figure \ref{fig:rheo} following \cite{reese1998theory}.  The left branch describes the mechanics of the nematic domains and the entropic elasticity of the rubber chain, and largely follows the treatment of Lee {\it et al.} \cite{lee_2023}.  The right branch adds viscoelasticity \cite{reese1998theory}.  We introduce a spontaneous stretch $F^s$, a network elastic deformation $F^n$, a viscous deformation $F^v$ and an elastic deformation $F^e$ that follow
\begin{equation} \label{eq:F2}
    F = F^sF^n = F^eF^v.
\end{equation}

The spontaneous stretch $F^s$ reflects the alignment of the domains in the polydomain material.  We follow Lee {\it et al.} \cite{lee_2023} and postulate that it depends on three internal variables -- scalars $\Lambda$ and $\Delta$, and a rotation $Q$ -- as follows: 
\begin{equation} \label{eq:G} 
    F^s = F^s(\Lambda,\Delta, Q) = G^{1/2}, \quad
    G=QG_0Q^T, \quad     
    G_0 = \begin{bmatrix}
    \Lambda^2 &0 &0\\
0 &\Delta^2/\Lambda^2 &0\\
0 &0 &1/\Delta^2
\end{bmatrix}.
\end{equation}
The internal variable $\Lambda$ describes the degree of nematic order, while the internal variable $\Delta$ describes the degree of planar order.  They are subject to the constraints,
\begin{equation} \label{eq:ag}
(\Lambda, \Delta) \in {\mathcal A}_G = \{(\Lambda, \Delta): \Delta \le r^{1/6}, \Delta \le \Lambda^2, \Delta \ge \sqrt{\Lambda}\}.
\end{equation}
 The values of the internal variables for some special states are shown in Table \ref{tab:iv}.  $Q$ denotes the orientation of spontaneous stretch.

\begin{table}
\centering
\caption{The internal variables and spontaneous stretch \label{tab:iv}}
\begin{tabular}{lcc}
\hline
Domain alignment & $\{\Lambda, \Delta\}$ & $G_0^{1/2}$\\
\hline
Random or isotropic & ${1,1}$ & $\begin{pmatrix}
1 &0 &0\\
0 &1 &0\\
0 &0 &1
\end{pmatrix}$ \\
\hline
Fully aligned & $\{r^{1/3},r^{1/6}\}$ & $\begin{pmatrix}
r^{1/3}&0 &0\\
0 &r^{-1/6} &0\\
0 &0 & r^{-1/6}
\end{pmatrix}$ \\
\hline
Random in-plane or equi-biaxial & $\{r^{1/12},r^{1/6}\}$ & $\begin{pmatrix}
r^{1/12}&0 &0\\
0 &r^{1/12} &0\\
0 &0 & r^{-1/6}
\end{pmatrix}$ \\
\hline
\end{tabular}
\end{table}

For future use, we denote the spatial velocity gradient as $L = \dot{F}F^{-1}$, and its viscous contribution as $L^v = \dot{F^v}(F^v)^{-1}$.  It follows that 
\begin{equation}
    L = \dot{F^e}(F^e)^{-1} + F^e\dot{F^v}(F^v)^{-1}(F^e)^{-1}.
\end{equation}

\subsection{Balance Laws}
The balance laws are standard,
\begin{align}
& \rho \ddot y = \nabla \cdot P + b &\text{(linear momentum)}, \\
& F P^T = P F^T &\text{(angular momentum)}, \\
& \dot{U} - s +\nabla \cdot h -P\cdot \dot{F} = 0 & \text{(energy)},\\
&  \dot{\eta}-s+\nabla \cdot (\frac{h}{T}) \geq 0 & \text{(Clausius-Duhem inequality)}
\end{align}
 in local Lagrangian form where $\rho$ is the density, $P$ the Piola-Kirchhoff stress, $b$ the body force, $U$ the internal energy density, $s$ the heat source, $h$ the heat flux and $\eta$ the entropy density (note that all densities are per unit reference volume). 
We introduce a free energy density $W = U - T\eta$, and use the balance of energy in the Clausius-Duhem inequality to rewrite it as
\begin{equation} \label{eq:diss}
    P\cdot \dot{F} -\dot{W} - \eta\dot{T} - \frac{1}{T}h \cdot \nabla T \geq 0. 
\end{equation}

\subsection{Constitutive framework}

We postulate that 
\begin{equation}
\begin{cases}
    W = \hat{W}(F,F^e,\Lambda,\Delta,Q,T)\\
    P = \hat{P^e}(F,F^e,\Lambda,\Delta,Q,T)\\
    \eta = \hat{\eta}(F,F^e,\Lambda,\Delta, Q,T)\\
    h=\hat{h}(\nabla T, T).
\end{cases}
\end{equation}

Substituting it into the inequality (\ref{eq:diss}) yields
\begin{equation}
\begin{aligned}
    &\Biggl( \hat{P^e}(F,F^e,\Lambda,\Delta, Q, T) - \frac{\partial \hat{W}}{\partial F}  - \frac{\partial \hat{W}}{\partial F^e}(F^v)^{-T} \Biggl) \cdot \dot{F}
    + \frac{\partial \hat{W}}{\partial F^e} \cdot F^e L^v\\
    & \hspace{0.75in} - \frac{\partial \hat{W}}{\partial \Lambda} \dot{\Lambda}
    - \frac{\partial \hat{W}}{\partial \Delta} \dot{\Delta}
    - \frac{\partial \hat{W}}{\partial Q} \dot{Q} 
    - \left( \frac{\partial \hat{W}}{\partial T} + \eta \right) \dot{T}  
    - \frac{1}{T}h \cdot \nabla T \geq 0.
\end{aligned}
\end{equation}
We expect that this inequality holds for all processes consistent with the balance of momenta and energy.  Therefore, arguing as in Coleman and Noll \cite{coleman_1963}, we conclude
\begin{equation} \label{eq:const2}
    \hat{P^e}= \frac{\partial \hat{W}}{\partial F} + \frac{\partial \hat{W}}{\partial F^e}(F^v)^{-T}, \qquad \eta = - \frac{\partial W}{\partial T},  \qquad h \cdot \nabla T \leq 0,
\end{equation}
The inequality now reduces to 
\begin{equation}
    (F^e)^T \frac{\partial \hat{W}}{\partial F^e} \cdot L^v
    - \frac{\partial \hat{W}}{\partial \Lambda} \dot{\Lambda}
    - \frac{\partial \hat{W}}{\partial \Delta} \dot{\Delta}
    - \frac{\partial \hat{W}}{\partial Q} \dot{Q} \geq 0.
\end{equation}
The frame-indifference of $\hat{W}$ means that $ (F^e)^T \partial \hat{W}/\partial F^e$ is symmetric, and so we can replace $L^v$ with its symmetric part or strain-rate $D^v = \frac{1}{2} (L^v + (L^v)^T)$ in the inequality.  We conclude
\begin{equation}
    (F^e)^T \frac{\partial \hat{W}}{\partial F^e} \cdot D^v
    - \frac{\partial \hat{W}}{\partial \Lambda} \dot{\Lambda}
    - \frac{\partial \hat{W}}{\partial \Delta} \dot{\Delta}
    - \frac{\partial \hat{W}}{\partial Q} \dot{Q} \geq 0.
\end{equation}

We note that each term is a product of an energetic quantity and a rate quantity, and therefore use it to identify the driving forces 
\begin{equation}
    \begin{cases}
         d_v =  (F^e)^T\frac{\partial \hat{W}}{\partial F^e} \qquad &\text{for $L^v$,}\\
         d_\Lambda =  - \frac{\partial \hat{W}}{\partial \Lambda} \qquad & \text{for $\dot{\Lambda}$,} \\
         d_\Delta = - \frac{\partial \hat{W}}{\partial \Delta} \qquad & \text{for $\dot{\Delta}$,} \\
         d_Q = \text{skew}\left(- \frac{\partial{\hat{W}}}{\partial Q}Q^T  \right) \qquad & \text{for $A$ \text{, where} $\dot{Q}=AQ$.}
    \end{cases}
\end{equation}
that determine the evolution of the strain rate $D^v$, state variables $\Lambda, \Delta$ and microstructure orientation $Q$ respectively.  Note that $Q \in SO(3)$, and hence $\dot{Q}=AQ$ for some $A$ skew, and it is natural to only look at the skew part of the final term.  Also, note that $d_v$ is the Cauchy stress associated with the network elasticity.  We postulate
\begin{align} \label{eq:kin}
\begin{cases}
D^v = \hat{D}^v(d_v),\\
\dot{\Lambda} = \hat{k}_\Lambda (d_\Lambda),\\
\dot{\Delta} = \hat{k}_\Lambda (d_\Delta),\\
\dot{Q} = \hat{A}(d_Q).
\end{cases}
\end{align}

In summary, the constitutive relations are specified as $\hat{W}, \hat{L}^v, \hat{k}_\Lambda, \hat{k}_\Delta, \hat{A}$ with the remaining quantities are obtained from (\ref{eq:const2}).

Before we describe specific constitutive relations, we note that the energy balance may be written as 
\begin{equation} \label{eq:en}
    c\dot{T}  = s-\nabla \cdot h + d_\Lambda \dot{\Lambda} + d_\Delta \dot{\Delta} +d_Q \dot{Q} +d_v \cdot L^v.
\end{equation}
where 
\begin{equation}
c = -\frac{\partial^2 W}{\partial T^2} \, \, T
\end{equation}
is the specific heat.

\subsection{Specific constitutive relations}

We begin with the energy, and assume
\begin{equation}
\begin{aligned}
\hat{W}(F,F^e,\Lambda,\Delta,Q,T) &= \hat{W}^n(F,\Lambda,\Delta,Q,T) + \hat{W}^e (F^e,T) + W^t(T) 
\end{aligned}
\end{equation}
where $W^n$ describes the nematic elasticity, $W^e$ the network elasticity that contributes to the viscoelasticity and $W^t$ is the purely thermal energy.
We further assume that  
\begin{equation}
\hat{W}^n(F,\Lambda,\Delta,Q,T) = \hat{W}^{ne}(F^T Q G_0^{-1}(\Lambda,\Delta) Q^T F,T) + \hat{W}^{nn}(\Lambda,\Delta,T).
\end{equation}
$\hat{W}^{ne}$ accounts for the entropic elasticity of the polymer chains averaged over the domains, and $W^{nn}$ is the stored energy that results when the domains are forced out of their as-synthesized isotropic state.  
We assume that $\hat{W}$ is isotropic (since the I-PLCE is synthesized in the isotropic state).  This means that $\hat{W}^{ne}$ depends on the principal invariants of $F^T Q G_0^{-1} Q^T F$.  Further, it is observed that the microstructure can reorient extremely quickly.  In other words, $d_Q = 0$, and we can minimize out $Q$ from the energy and replace $\hat{W}^{ne}$ with 
\begin{equation}
\tilde{W}^{ne} (F,\Lambda,\Delta,T) = \min_Q \ \hat{W}^{ne}(F,\Lambda,\Delta,Q,T).
\end{equation}
We take
\begin{equation}
\tilde{W}^{ne} (F,\Lambda,\Delta,T) = \frac{\mu_1}{2} \left( J^{-2/3} \left(\frac{\lambda_1^2}{\Lambda^2} + \frac{\lambda_2^2\Lambda^2}{\Delta^2} + \lambda_3^2 \Delta^2\right) - 3 \right) + \frac{\kappa_1}{2}  \left(\frac{J^2-1}{2}-\ln J \right)
\end{equation}
where $\lambda_1 \ge \lambda_2 \ge \lambda_3$ are the ordered principal values of $F$ and $J= \text{det} \, F = \lambda_1  \lambda_2  \lambda_3$ by extending the incompressible formulation of Lee {\it et al.} \cite{lee_2023} to the compressible case.  We also assume 
\begin{equation}
 \hat{W}^{nn}(\Lambda,\Delta,T) 
 = C\frac{\Delta-1}{(r^{1/6}-\Delta)^k} \ ,
 \end{equation}
 \begin{equation}
        \hat{W}^e = \frac{\mu_2}{2} \left(J_e^{-2/3}\text{tr}(F^{eT}F^e)-3 \right) + \frac{\kappa_2}{2} \left(\frac{J_e^2-1}{2}-\ln J_e \right)
\end{equation}
and
\begin{equation}
        \hat{W}^t = - c T \ln T.
\end{equation}

Turning now to the kinetics or evolution (\ref{eq:kin}), we assume linear kinetics for the state variables:
\begin{align}
\dot{\Lambda} = \alpha_\Lambda d_\Lambda, \quad 
\dot{\Delta} = \alpha_\Delta d_\Delta
\end{align}
for kinetic coefficients $\alpha_\Lambda, \alpha_\Delta$.  
Finally, we follow the practice in viscoelasticity and rewrite the first of (\ref{eq:kin}) as $d_v = \hat{d}_v (D^v)$.  We also decompose the strain rate into hydrostatic and deviatoric parts, and assume a nonlinear relation for the deviatoric evolution and linear relation for the hydrostatic viscosity:
\begin{equation}
d_v = \nu_h (\text{tr} \ D^v) I + \nu_d ( || D^{v,\text{dev}} || ) D^{v,\text{dev}} 
\end{equation}
where $I$ is the identity,  $D^{v,\text{dev}} = D^v - \frac{1}{3} (\text{tr} \ D^v) I$ is the deviatoric part of $D^v$, $\nu_h$ is the volumetric viscosity, $\nu_d (\dot \gamma)$ is the rate-dependent viscosity (it is shear hardening if it is increasing and shear weakening if it is decreasing) that satisfies $\nu_d (0) = 0, \nu_d \ge 0$.   We specifically take $\nu_d$ to be in the Cross form \cite{cross}
\begin{equation}
\nu_d (\dot{\gamma}) = \nu_\infty + \frac{\nu_0 - \nu_d}{1 + (\dot{\gamma}/\dot{\gamma}_0)^m}
\end{equation}
for constants $\nu_0$ (creeping viscosity), $\nu_\infty$ (high rate viscosity), $\dot{\gamma}_0$ (reference shear rate) and $m$ (viscosity exponent).

\section{Results and Discussion}

\subsection{Tension}
Unaligned, isotropic-genesis, polydomain liquid crystal elastomers (I-PLCEs) are tested at three nominal strain rates of 0.01 s$^{-1}$, 100 s$^{-1}$, and 300 s$^{-1}$. The samples used in both 100 s$^{-1}$ and 250 s$^{-1}$ tests come from two different batches, but we have verified that the batch-to-batch variation is small compared to the rate effects. 

\paragraph{Quasistatic tests}

\begin{figure}
\centering
\includegraphics[width=0.50\textwidth]{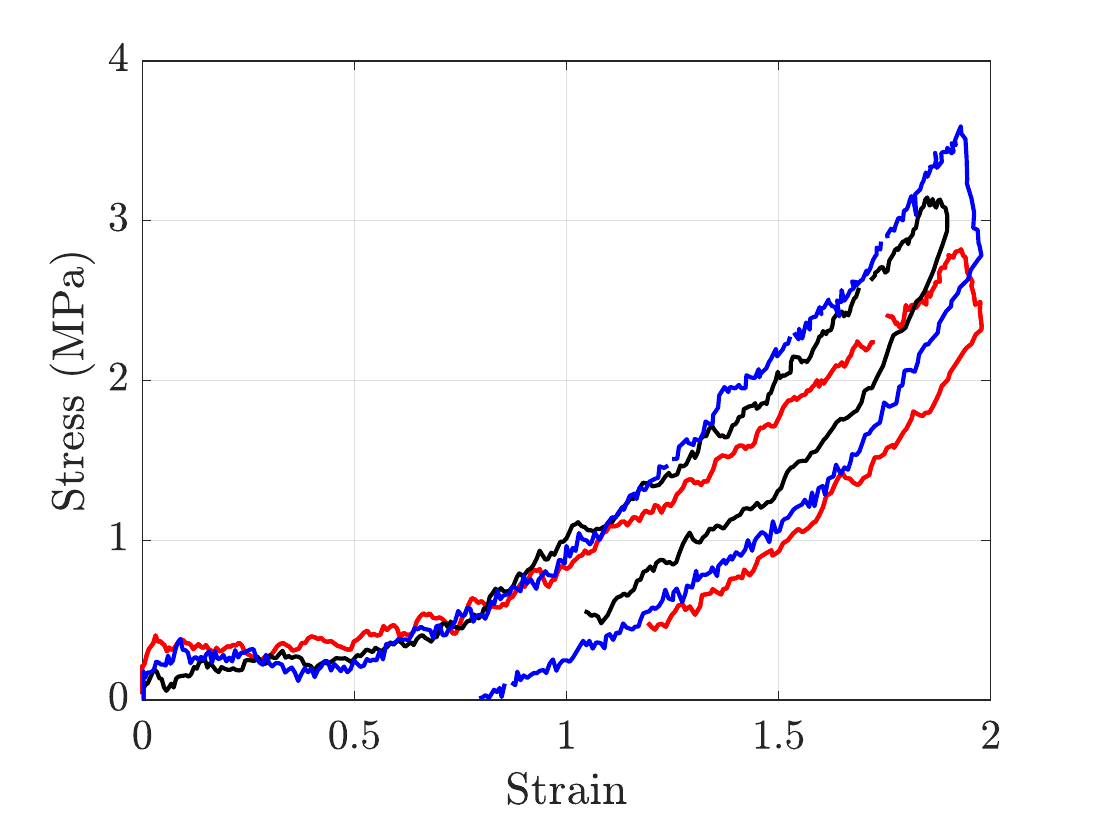}
\caption{Behavior of I-PLCE in tension at quasistatic strain rate of $\dot{\varepsilon} = 10^{-2}$ s$^{-1}$.}
\label{fig:qs}
\end{figure}

Figure \ref{fig:qs} shows the stress-strain behavior of the I-PLCE at a strain rate of $\dot{\varepsilon} = 10^{-2}$ s$^{-1}$.  We observe the characteristic soft behavior with a stress plateau of about 0.3 MPa.

\paragraph{Tests at  100 s$^{-1}$}

\begin{figure}
\centering
\includegraphics[width=6in]{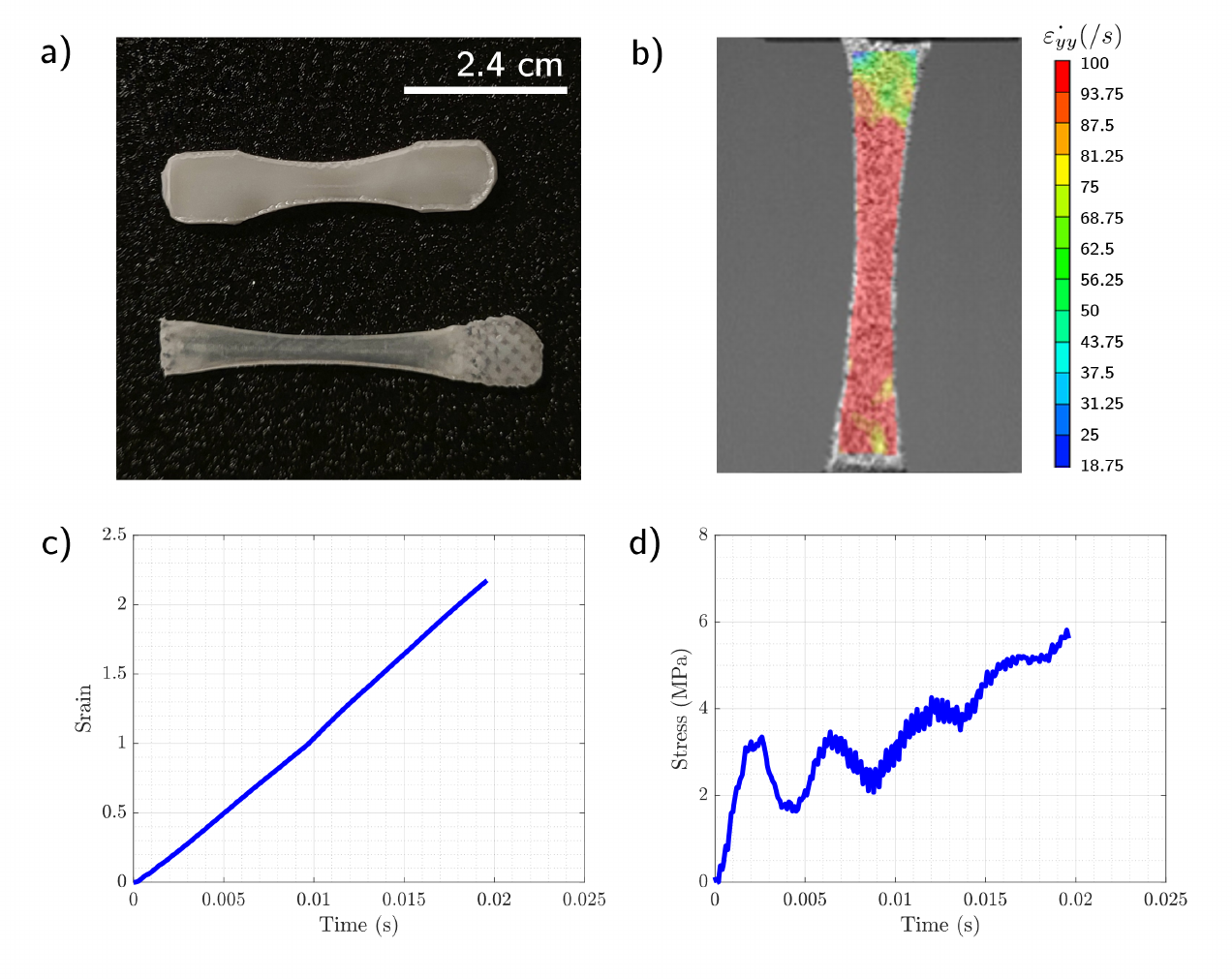}
\caption{Typical results of a test at  3.4 m/s striker velocity or 100 s$^{-1}$ nominal strain rate. (a) Specimen before and after the test. (b) Snapshot of the specimen during the test superposed with the DIC strain rate in the stretching direction. s(c) Green-Lagrange strain as a function of time. (d) Piola-Kirchhoff or engineering stress as a function of time} 
\label{fig:typical}
\end{figure}

\begin{figure}
	\centering
	\includegraphics[width=1.0\textwidth]{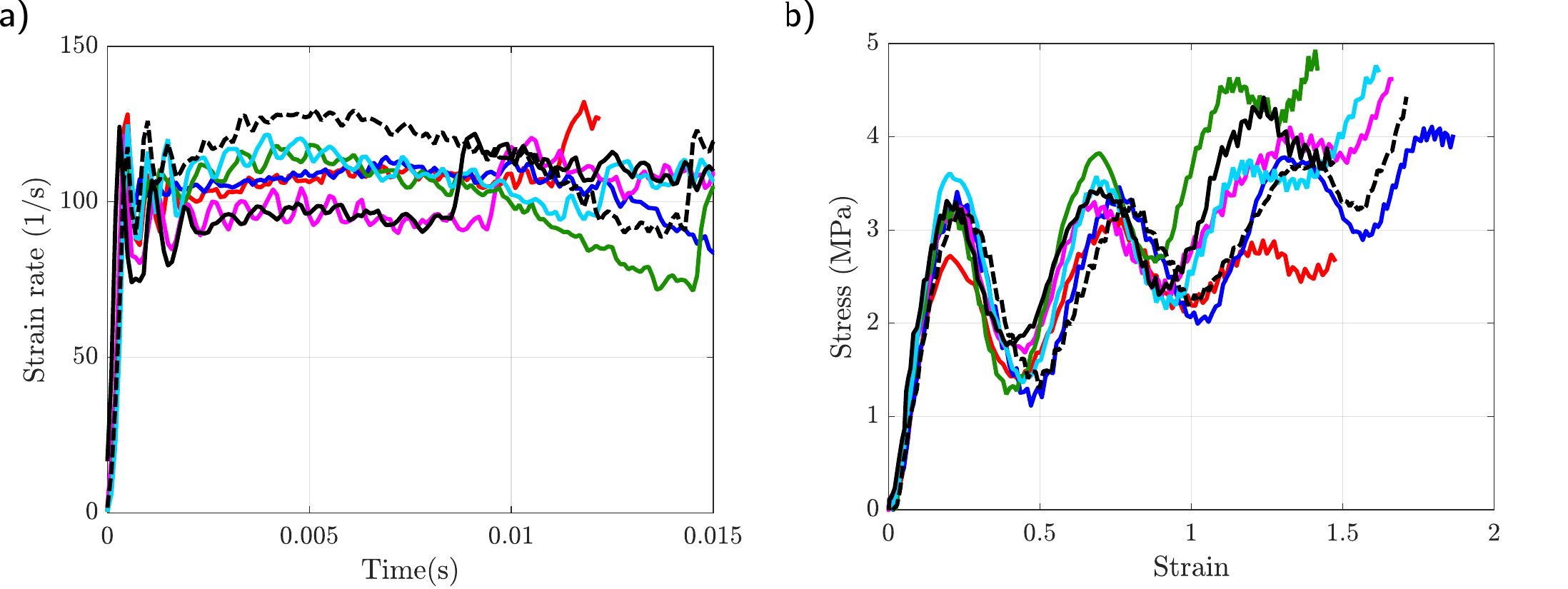}
	\caption{The observed response of the I-PLCE at a nominal strain rate of 100 s$^{-1}$. (a) Strain rate evolution in time and (b) Stress and strain response.}
	\label{fig:srss100}
\end{figure}

We obtain a striker velocity of 3.4 m/s with a drop height of 0.7  m, and this leads to a nominal strain rate of 100 s$^{-1}$ as we see later.  Typical results are shown in Figure \ref{fig:typical}.  The specimen before (top) and after (bottom) the test is shown in Figure \ref{fig:typical}(a). Note that the test proceeds until specimen failure, which typically occurs at the grip. While this indicates that failure is influenced by local stress concentrations at the clamping region, the stress-strain response of large strains up to failure is reproducible and representative of the bulk material behavior. A typical snapshot superposed with the DIC strain field is shown in Figure \ref{fig:typical}(b), while a typical observation of the Green-Lagrange strain and Piola-Kirchhoff engineering stress as a function of time is shown in Figures \ref{fig:typical}(c) and (d) respectively.  We observe oscillations in the stress response with a frequency of about 4 milliseconds. These oscillations are present across all our experiments, and as discussed extensively in our previous work \cite{nieto2025new}, are artifacts created by the interaction between flexural waves in the steel bar and load sensors.  So we disregard these oscillations as measurement artifacts and focus on the underlying stress-strain trends.

The strain rate as a function of time, as well as the stress-strain behavior, over a number of tests is shown in Figures \ref{fig:srss100}(a) and (b).  We observe that the strain rate is almost constant at a value of 100 s$^{-1}$.  Therefore, we regard these tests as tests at a nominal strain rate of 100 s$^{-1}$.   In general, the stress-strain response at this rate is much stiffer than that in the quasi-static regime shown in Figure \ref{fig:qs}. For example, the peak stress at 100 s$^{-1}$ reaches 3 MPa, compared to 0.3 MPa at 0.01 s$^{-1}$.

The soft behavior corresponding to polydomain to monodomain transition is not immediately evident in Figure \ref{fig:srss100}: the response is initially elastic but then transitions to a flow-type behavior (with the oscillations discussed earlier).  However, we observe that the specimen changes from opaque to translucent during the tests (see Figure \ref{fig:typical}(a)), showing that there is indeed alignment of the domains or mesogens towards the loading direction.  So we conclude that the soft behavior is present, but masked by the rate effects, in the macroscopic stress-strain measurements.

Finally, there is some variability in the observations of strain rate and stress-strain response due to variability in batches and tests. The standard deviation of strain rate is 9.5 s$^{-1}$, and the strain standard deviation is 0.065 at 1.4 strain. The stress standard deviation is 0.6 MPa at 3.8 MPa. However, we do not observe a clear correlation between higher nominal strain rates and higher stress: compare the two tests marked in black in the figure: the test indicated in a solid line has a lower strain rate, but a higher stress value.

\paragraph{Tests at 300 s$^{-1}$}

\begin{figure}
    \centering
    \includegraphics[width=1.0\textwidth]{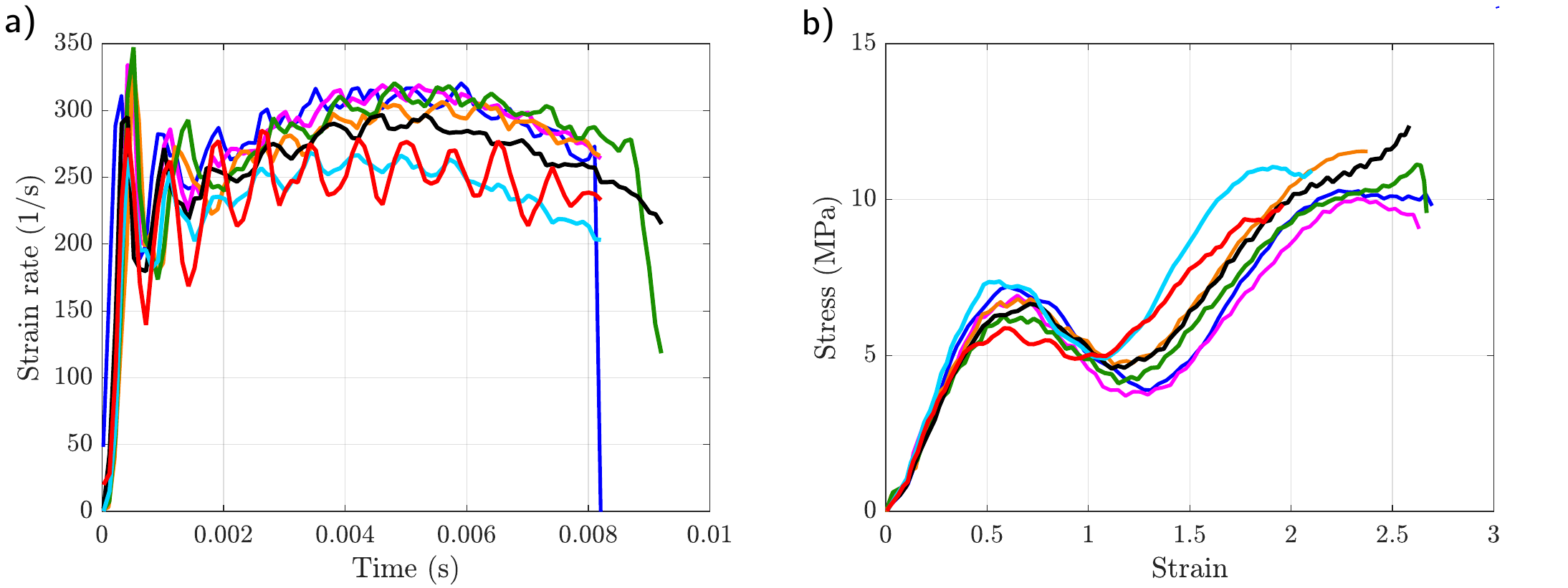}
    \caption{The observed response of the I-PLCE at a nominal strain rate of 250-300 s$^{-1}$. (a) Strain rate evolution in time and (b) Stress and strain response.}
    \label{fig:srss300}
\end{figure}

We obtain a striker velocity of 5.4 m/s with a drop height of 1.6 m.  The strain rate as a function of time, as well as the stress-strain behavior, over a number of tests is shown in Figure \ref{fig:srss300}(a) and (b).  We observe that the strain rate is not constant over time (in contrast to the previous set of tests at a striker velocity of 3.4 m/s).  Instead they start off at around 250 s$^{-1}$ and reach a peak of about 300 s$^{-1}$ before falling back to around  250 s$^{-1}$.  We label these tests with the peak strain rate of 300 s$^{-1}$.  

We observe that we have an initial elastic behavior until about a stress of 7 MPa which is significantly higher than the 3 MPa observed in tests with striker velocity 3.4 m/s, and the 0.3 MPa in quasistatic tests.  We have a flow-like behavior after the initial elastic behavior with a hardening rate that is higher than in the previous tests with striker velocity 3.4 m/s.  We do not see any identifiable soft behavior in the stress-strain curve. We also observe that the specimen shows little change in transparency, unlike the tests at 100 s$^{-1}$.  There is some variability due to variations in batches and tests. The standard deviation of the strain rate is 20 s$^{-1}$. At a strain of 2, the standard deviation of the strain is 0.3 with a stress standard deviation of 0.5 MPa.

\paragraph{Discussion}

Comparing across all experiments, we conclude that the stress response is strain rate dependent, with the peak stress increasing with increasing strain rate. These are consistent with the finding of \cite{linares} and \cite{azoug} who tested LCEs at much small rates ($10^{-4}$ s$^{-1}$ to 1 s$^{-1}$).  We also conclude that the plateau behavior of the PMT regime, as well as the onset of this regime, are affected by the strain rate. The limited softness at high strain rates can be attributed to the timescale associated with domain reorientation, again consistent with previous observations at lower rates \cite{hotta_terentjev, azoug, linares, clarke_terentjev}.

\subsection{Compression}

The behavior of I-PLCE pads (\ref{sec:MTS}) in compression is studied at three different strain rates of 0.01 s$^{-1}$, 8000 s$^{-1}$, and 19500 s$^{-1}$.

\begin{figure}
\centering
    \includegraphics[width=3in]{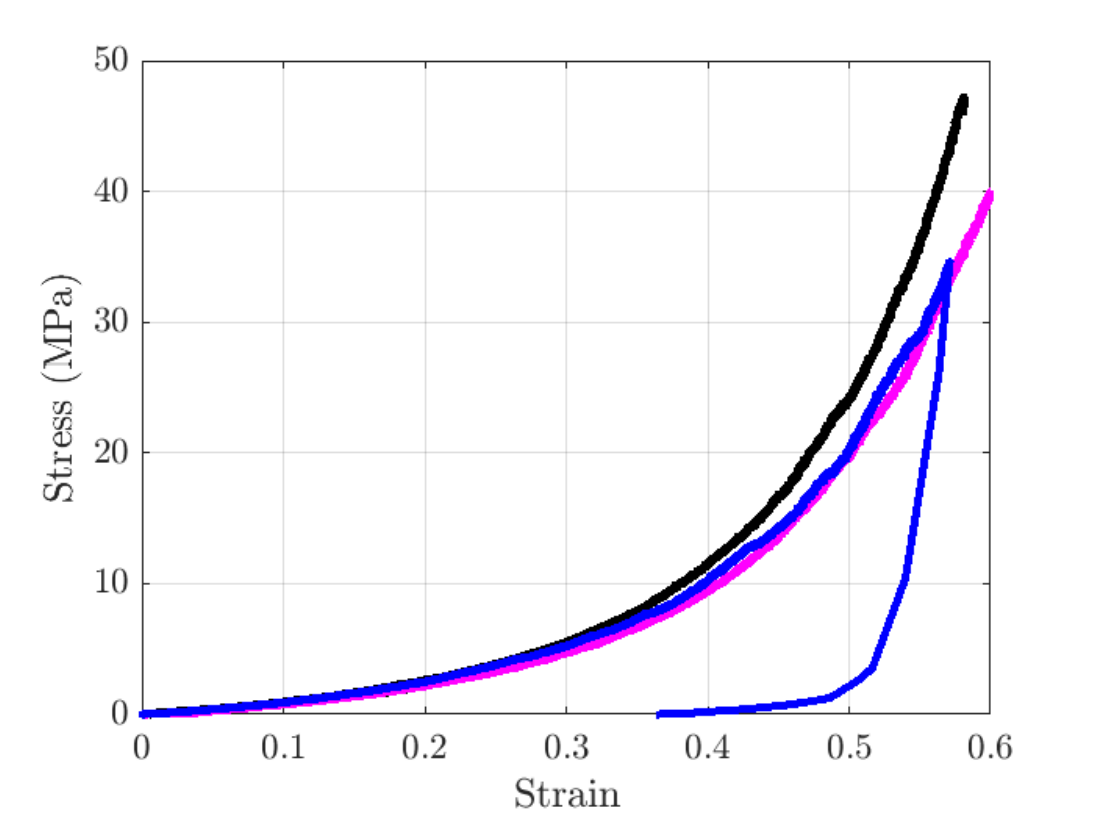}
    \caption{Behavior of LCE pads in compression at quasistatic strain rate of $\dot{\varepsilon} = 10^{-2}$ s$^{-1}$.}
    \label{fig:compqs}
\end{figure}

\paragraph{Quasistatic tests} To avoid buckling under compression, we test LCE pads with a 10 mm diameter and 3 mm thickness. The quasistatic response of I-PLCE pads at a strain rate of 10$^{-2}$ s$^{-1}$ is shown in Figure \ref{fig:compqs}.  The stress increases nonlinearly with strain, reaching up to 35 MPa at a strain of 0.55. We do not see any soft behavior, consistent with the observations of Saed {\it et al.} \cite{saed_impact}.

\paragraph{Tests at 8000 s$^{-1}$} 
\begin{figure}
\centering
    \includegraphics[width=1.0\textwidth]{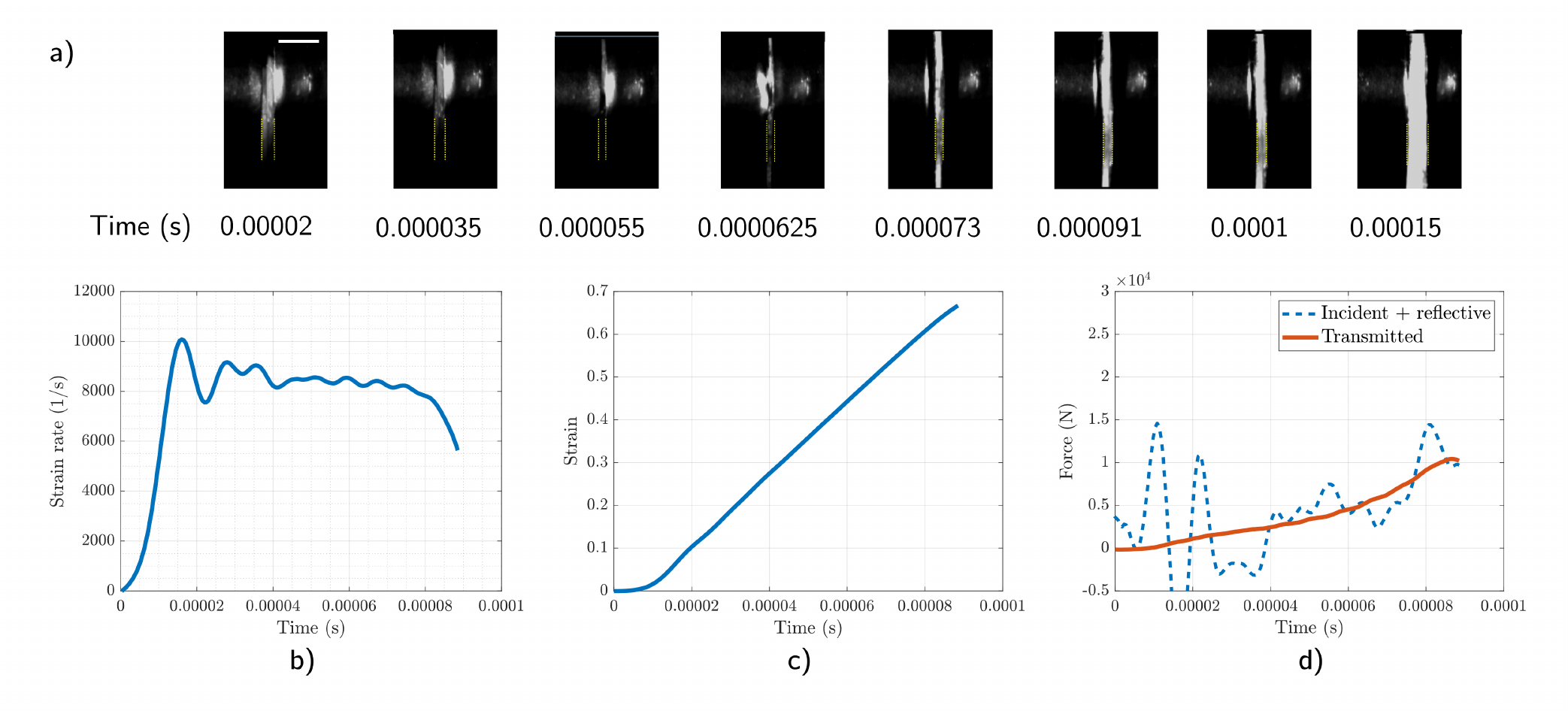}
    \caption{Typical results of a test at a nominal strain rate of 8000 s$^{-1}$.  (a)  Snapshots showing the specimen being compressed between the incident and the transmitted bar (scale bar is 3 mm).  (b) Axial strain as a function of time.  (c) Strain rate as a function of time. (d) Force acting on the specimen due to the incident and reflected pulse, as well as the transmitted pulse, as a function of time.}
    \label{fig:typicalc}
\end{figure}

We test LCE pads with a 6.7 mm diameter and 3 mm thickness. Typical results of a test performed at a nominal strain rate of 8000 s$^{-1}$ are shown in Figure \ref{fig:typicalc}.  Figure \ref{fig:typicalc}(a) shows a series of snapshots of the specimen being compressed between the incident and the transmitted bar of the SHPB apparatus.  We observe that the specimen is compressed as the wave arrives and until the incident bar separates from the specimen.  There are, in fact, several impacts (see video in Supplementary Materials) due to the reverberations of the incident bar, and the material begins to disintegrate (fragment).

We only consider the measurements from the time between the arrival of the incident wave and the first separation of the incident bar.   Figure \ref{fig:typicalc}(b) shows the axial strain inferred from the measurements in the incident and transmitted bars as a function of time. Figure \ref{fig:typicalc}(c) shows the forces on the specimen at the specimen-bar interface (input and output bar) as a function of time. We see that the two forces are balanced, indicating that the specimen is in dynamic equilibrium.

\begin{figure}
	\centering
	\includegraphics[width=6in]{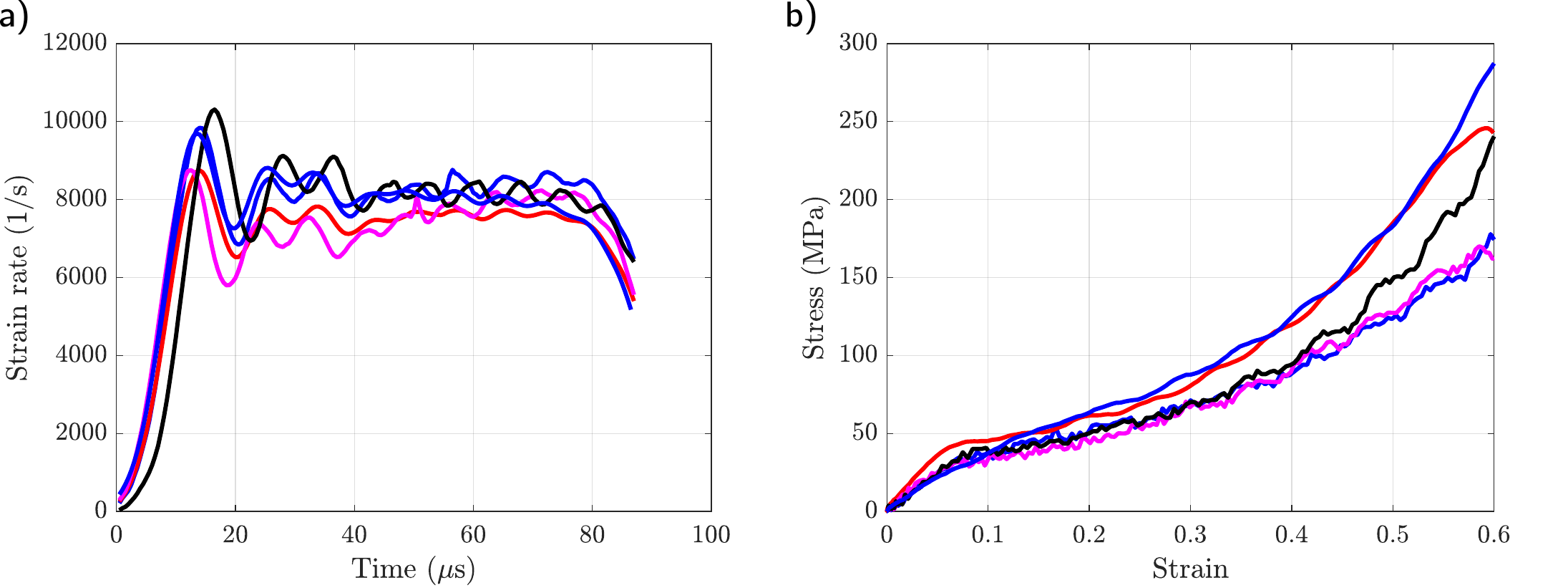}
	\caption{The observed response of the I-PLCE in compression at a nominal strain rate of 8000 s$^{-1}$. (a) Strain rate evolution in time and (b) Stress and strain response.  }
	\label{fig:srss8000}
\end{figure}

The strain rate as a function of time, as well as the stress-strain behavior, over a number of tests is shown in Figure \ref{fig:srss8000}(a) and (b).  We observe that the strain rate is almost constant during the test.    We also observe that the stress increases in a nonlinear fashion to reach approximately 100 MPa at a strain of 0.5, and these values are significantly higher than those in the quasistatic test.  We again do not observe any evident soft behavior.

There is again some variability in the strain rate and stress-strain response across batches. The strain rate has a standard deviation of approximately 50 s$^{-1}$. At a strain of 0.1, the standard deviation of the strain is 0.013, and the stress is 4.5 MPa. At a strain of 0.25, the standard deviation of the strain is 0.03 and the stress is 13 MPa.

\paragraph{Tests at 19500 s$^{-1}$} 
\begin{figure}
	\centering
	\includegraphics[width=6in]{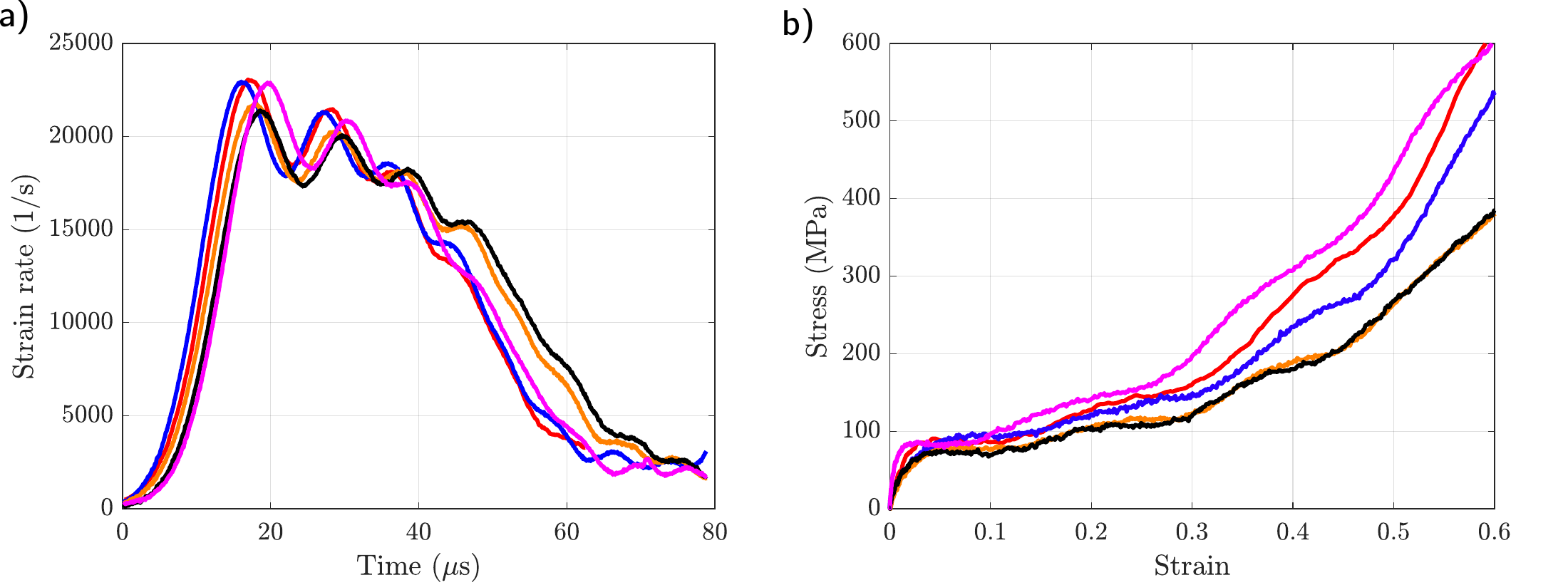}
	\caption{The observed response of the I-PLCE in compression at a nominal strain rate of 19500 s$^{-1}$. (a) Strain rate evolution in time and (b) Stress and strain response.}
	\label{fig:srss18000}
\end{figure}

We test LCE pads with a 6.7 mm diameter and 0.6 mm thickness. The tests are repeated with a higher striker velocity.  As before, there are several impacts, and the samples disintegrate. Indeed, the specimen disintegrates into small charred fragments, and this is accompanied by a strong smell characteristic of toluene. We deduce that the generation of heat during repeated impact, as well as the friction between the specimen and the bars ignite the highly flammable toluene (autoignition temperature of pure toluene is 536$^\circ$C \cite{SigmaAldrich_Toluene_244511}).  

We only use data from before the first separation.  The strain rate as a function of time, as well as the stress-strain behavior, over a number of tests is shown in Figure \ref{fig:srss18000}(a) and (b).  We observe that the strain rate is not constant over time (in contrast to the previous set of tests at 8000 s$^{-1}$).  The stress is even higher than that at 8000 s$^{-1}$, reaching 175 MPa at 0.5 strain. Again, all specimens do not exhibit soft behavior.

As expected, there is variability across batches. The strain rate has a standard deviation of approximately 2000 s$^{-1}$. At a strain of 0.1, the standard deviation of the strain is 0.03, and the stress is 26 MPa. At a strain of 0.25, the standard deviation of the strain is 0.03 and the stress is 40 MPa.

\paragraph{Discussion}

We do not see any visible soft behavior in any of our tests, including the quasi-static test, consistent with previous observations  \cite{saed_impact}. Further, since the specimens disintegrated on impact, it is not possible to visually examine the opacity change in the specimen that can indicate any domain reorientation.  Furthermore, we recall that the soft behavior in compression is expected to be significantly smaller than in tension: in the ideal case, we expect the plateau strain to be $1-r^{1/12}$ in compression as compared to $r^{1/3}-1$ in tension, where $r>1$ is the anisotropy parameter.  This can be further reduced due to lateral confinement.

The contact conditions between the specimens and the bars can affect the results significantly.  We applied petroleum jelly to reduce the friction.  Still, the geometry of our specimens (the thickness of the specimen is significantly smaller than its diameter) and the tackiness of LCEs make it inevitable that there is some slip and friction.  The images shown in Figure \ref{fig:typicalc} suggest the presence of slip, but the images do not provide sufficient resolution to quantify the amount of slip.  The friction contributes to lateral confinement, as does inertia in the pad geometry.  

The friction, the geometry, the dissipation in the material due to domain evolution, and viscosity all lead to a temperature increase. 
We can obtain an order of magnitude estimate of the temperature increase by considering the energy deposited during impact (the area below the stress-strain curve) and assuming adiabatic conditions and negligible elastic response so that all of the work is converted to heat.   Approximating the stress-strain behavior to be linear, we get that 
\begin{equation}
\Delta T =  \frac{1}{2c} \sigma_\text{max} \varepsilon_\text{max}
\end{equation}
where $c$ is the specific heat per unit volume, $\sigma_\text{max}$ the peak stress and $\varepsilon_\text{max}$ the highest strain.  The specific heat of LCEs is 1.57 J/g$^\circ$C and density is 1.25 g/cm$^3$, and so $c =1.93 \times 10^6$ Pa/$^\circ$C (J/m$^3$$^\circ$C).  Taking the peak stress to be 200 MPa and 500 MPa for 8000 and 19500 s$^{-1}$ cases respectively, and a maximum strain of 0.6, gives us a temperature rise of 30.3$^\circ$C and 75.8$^\circ$C for the 8000 and 19500 s$^{-1}$ cases.  This is, of course, an overestimation, since we assume that all work done is converted to heat.  Alternatively, we can use equation (\ref{eq:en}): using the parameters in Table \ref{tab:param} under the assumption of uniaxial compression, we obtain a temperature increase of 26 and 40 $^\circ$C temperature rise in the 8000 and 19500 s$^{-1}$ cases.  These, by themselves, are not high enough to lead to the self-ignition of toluene.  However, these ignore friction that would add to the temperature increase.  Further, there are multiple impacts, fragmentation of the specimen, and possible hot spots due to heterogeneity or collapse of a microvoid.  In fact, a single hot spot reaching the autoignition temperature of Toluene could be sufficient to initiate the whole reaction.  Finally, it is possible to have bar-to-bar contact in the thin 0.6 mm specimens used for the 19500 s$^{-1}$ case.

\subsection{Comparison between model and experiments}

\paragraph{Fitting the constitutive model}

\begin{table}
\centering
\begin{tabular}{ll}
\hline
Parameter & Value\\
\hline
Shear modulus (equilibrium), $\mu_1$ \hspace{0.5in} & $4.5$ MPa\\
Shear modulus (nonequilibrium), $\mu_2$ & $1.8$ MPa\\
Bulk modulus (equilibrium), $\kappa_1$ & $0.025$ GPa\\
Bulk modulus (nonequilibrium), $\kappa_2$ & $1.3$ GPa\\
LCE anisotropy parameter, $r$ & 6.0\\
Hardening coefficient, $C$ & 100 Pa\\
Hardening exponent $k$ & 2\\
Kinetic $\Delta$ coefficient, $\alpha_\Delta$ & $ 5.0 \times 10^3$ (Pa $\cdot$ s)$^{-1}$ \\
Kinetic $\Lambda$ coefficient, $\alpha_\Lambda$ & $ 5.0 \times 10^1$ (Pa $\cdot$ s)$^{-1}$\\
Hydrostatic Viscosity, $\nu_h$ & $1.0 \times 10^5 $ Pa $\cdot$ s\\
Deviatoric Viscosity, $\nu_d$:\\
Creeping viscosity,  $\nu_0$ & $1.0 \times 10^5 $ Pa $\cdot$ s\\
High rate viscosity, $\nu_\infty$ & $3.0 \times 10^6 $ Pa $\cdot$ s\\
Reference shear rate, $\dot{\gamma}_0$ & $3.71  \times 10^2$ s$^{-1}$\\
Viscosity exponent, $m$ & 2.9\\
\hline
\multicolumn{2}{p{4in}}{The deviatoric viscosity takes the following values at the various
uniaxial (normal) strain rates}\\
$\dot{\varepsilon}$ = 10$^{-2}$ s$^{-1}$, & $1.0 \times 10^5 $ Pa $\cdot$ s\\
$\dot{\varepsilon}$ = 100 s$^{-1}$ & $1.35 \times 10^5 $ Pa $\cdot$ s\\
$\dot{\varepsilon}$ = 300 s$^{-1}$ & $7.7 \times 10^5 $ Pa $\cdot$ s\\
$\dot{\varepsilon}$ = 8000 s$^{-1}$,  19500 s$^{-1}$ & $3.0 \times 10^6 $ Pa $\cdot$ s\\
\hline
\end{tabular}
\caption{Parameters used for fitting with tensile and compression experiments. }
\label{tab:param}
\end{table}

\begin{figure} 
\centering
\includegraphics[width=6.0in]{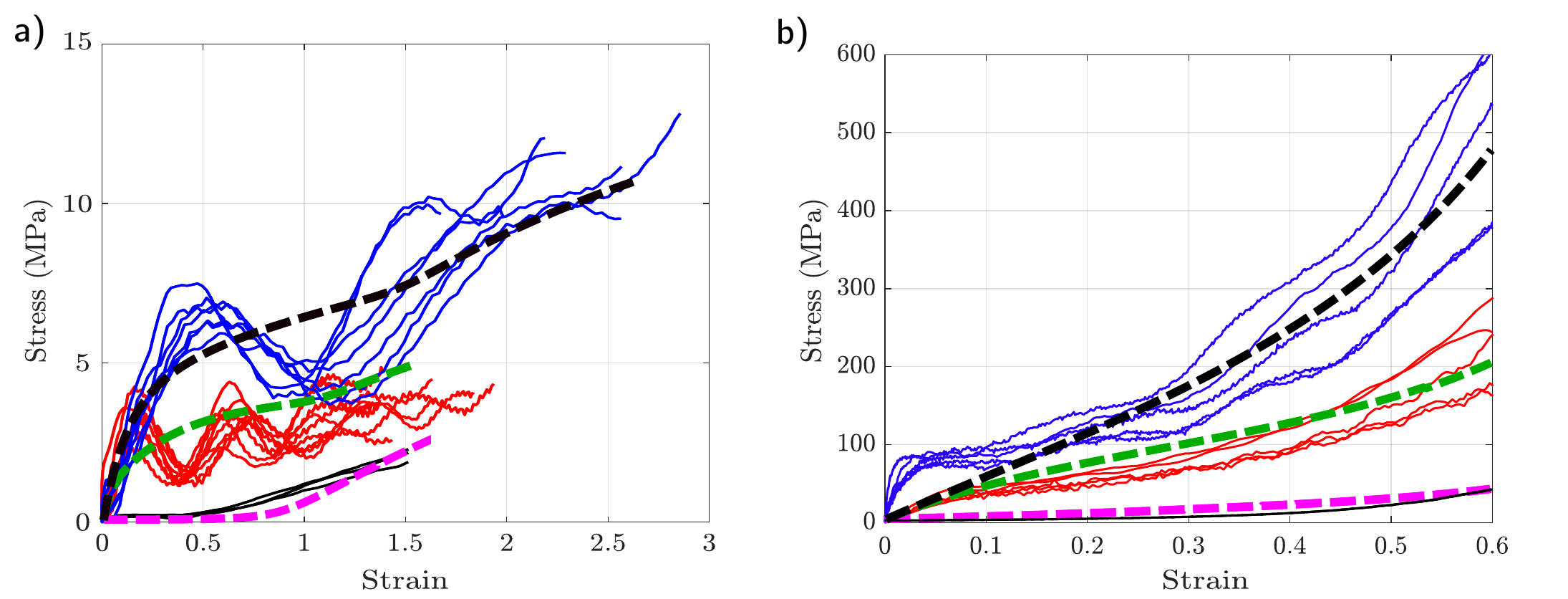}
\caption{Fit of the constitutive model presented in Section \ref{sec:model} to the experimental data in (a) tension and (b) compression. Parameters are given in Table \ref{tab:param}.}
\label{fig:fit}
\end{figure}

We examine the ability of the macroscopic model outlined in Section \ref{sec:model} to describe the experiments.   

We model the tensile tests by assuming a state of uniaxial stress and isothermal conditions in light of the geometry.  Specifically, we assume that the deformation gradient is the form $F = \text{diag}(\lambda_1, \lambda_2, \lambda_2)$, and that the Piola-Kirchhoff stress is $P = \text{diag}(\sigma,0,0)$.  This ignores the effect of the grips, but we believe that uniaxial stress is reasonable given the length of the specimen and our DIC observations.  Further, given the small thickness and large area, we believe that isothermal is a reasonable assumption.

We model the compressive tests by assuming a state of uniaxial strain in light of the small thickness of the specimen as well as the contact with the grips.  As discussed above, there is some slip between the specimen and the bar.  Therefore, a full model should take this as well as friction into account.  We can do so following full field inversion approaches, but this is beyond the scope of the paper.  So, we have to either choose a state of uniaxial stress or uniaxial strain to proceed.  We chose the latter.  Specifically we assume that $F = \text{diag}(\lambda, 1, 1)$.   We also assume isothermal conditions, and this is a significant assumption due to the temperature increase discussed above.

We fit the constitutive constants progressively.  We fit the anisotropy ratio, the equilibrium shear modulus, the hardening coefficient and exponent, and kinetic coefficients from quasi-static tension.  We then use the intermediate strain rate experiments to fit the non-equilibrium shear modulus and creeping viscosity.  Finally, we fit the high rate viscosity, the hydrostatic viscosity, and bulk moduli from the compression tests.  One could use emerging regression and optimization techniques, but we have not done so.  The resulting parameters are given in Table \ref{tab:param}, and the comparison between the experimental observations and model results with the chosen parameters are shown in Figure \ref{fig:fit}.   We emphasize that we use the same set of parameters in the comparison in Figure \ref{fig:fit}.   We note that the agreement is very good. 

We have a high degree of confidence in the modeling of the quasistatic and dynamic tension tests, and thus the ability of the model to describe processes  up to 10$^2$ s $^{-1}$ strain rates.  We have less confidence in the compressive due to contact, fragmentation, and temperature increase in the dynamic compression tests, and thus the ability of the model to describe processes in the 10$^3$ s $^{-1}$ strain rates.  Still, it is notable that we are able to describe the entire range of tests and conditions with one single set of model parameters.

\paragraph{Suppression of mesogen and domain reorientation at high rates}

\begin{figure} 
\centering
\includegraphics[width=6in]{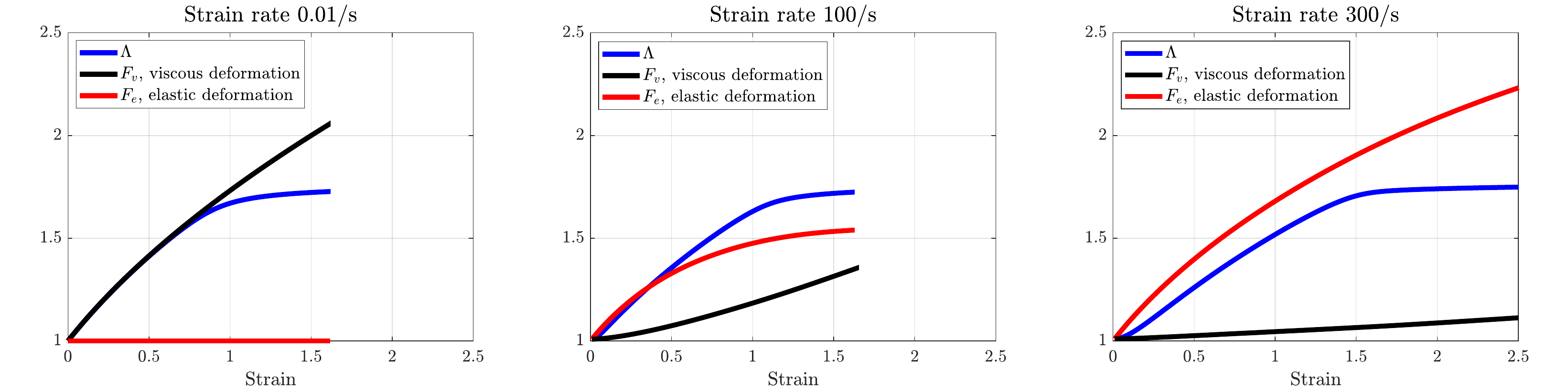}
\caption{The evolution of various internal variables in the model.  Parameters are given in Table \ref{tab:param}}
\label{fig:evol}
\end{figure}

We finally use the model to gather further insight into the behavior of the materials.  We do so by examining the evolution of the internal variable $\Lambda$ as well as the uniaxial elastic and viscous deformations (largest principal values of $F^e$ and $F^v$, respectively).  This is shown in Figure \ref{fig:evol} for the tensile tests.  At very small strain rates, we see that the internal variable $\Lambda$ evolves till close to the value of 1.82 (corresponding to $r^{1/3}$ for $r=6$) 
till an imposed strain of about 0.8.  We see very little elastic strain.  This means that much of the deformation is a result of the mesogen and domain reorientation, resulting in the soft behavior.  At higher strain rates, we observe that the evolution of the $\Lambda$ is retarded while the elastic strain increases rapidly.  This indicates that the viscosity masks the mesogen and domain reorientation as observed in the experiments.  In compression, the ability of the I-PLCE to undergo domain rotation is already reduced since it changes from isotropic to equi-biaxed.  Further, the strain rates are significantly higher, and so viscosity suppresses it further.

\section{Conclusion}

In this study, we probe the behavior of isotropic-genesis polydomain liquid crystal elastomers (I-PLCEs) over a wide range of strain rates: $10^{-2}$ to $3 \times 10^{2}$ s$^{-1}$ in tension and $10^{-2}$ to $19.5 \times 10^{3}$ s$^{-1}$ in compression. Tensile drop tower tests reveal a clear rate-dependent stiffening in the stress-strain response and a limitation of the classical soft elasticity at high rates. Compression split-Hopkinson pressure bar tests, similarly, show a rate-dependent stiffening and limited soft-behavior of I-PLCE pads. These observations indicate that at high strain rates, away from equilibrium, the rate-dependency of the entropic network relaxation is not trivial. We propose a thermodynamically-consistent model that incorporates both the nematic-order evolution as well as polymer viscosity, showing good agreement across all studied strain-rate regimes. Our model captures the coupling of mesogen kinetics and viscoelastic flow across various strain rates as demonstrated by its ability to fit the whole range of tests with a single set of parameters.  Therefore, it can serve as a predictive framework for designing I-PLCEs for various applications.

\section*{Acknowledgement}
We gratefully acknowledge the support of the National Science Foundation (2009289) and the Army Research Office (W911NF-22-1-0269).


\end{document}